\documentclass[prl,twocolumn,superscriptaddress,floatfix]{revtex4-1}
\usepackage{graphicx}
\usepackage{multirow}
\begin{document}
\title{Effect of a built-in electric field in asymmetric ferroelectric tunnel junctions}
\author{Yang Liu}
\email{liuyangphy52@gmail.com}
\affiliation{Multi-disciplinary Materials Research Center,
Frontier Institute of Science and Technology,
Xi'an Jiaotong University, Xi'an 710054, People's Republic of China}
\affiliation{Laboratoire Structures, Propri\'{e}t\'{e}s et Mod\'{e}lisation des Solides,
UMR 8580 CNRS-Ecole Centrale Paris, Grande Voie des Vignes, 92295 Ch\^{a}tenay-Malabry Cedex, France}
\author{Xiaojie Lou}
\affiliation{Multi-disciplinary Materials Research Center, Frontier Institute of Science and Technology,
Xi'an Jiaotong University, Xi'an 710054, People's Republic of China}
\author{Manuel Bibes}
\affiliation{Unit\'{e} Mixte de Physique CNRS/Thales, 1 Av. A. Fresnel, Campus de l'Ecole Polytechnique,
91767 Palaiseau and Universit\'{e} Paris-Sud, 91405 Orsay, France}
\author{Brahim Dkhil}
\affiliation{Laboratoire Structures, Propri\'{e}t\'{e}s et Mod\'{e}lisation des Solides,
UMR 8580 CNRS-Ecole Centrale Paris, Grande Voie des Vignes, 92295 Ch\^{a}tenay-Malabry Cedex, France}

\begin{abstract}
The contribution of a built-in electric field to ferroelectric phase transition in
asymmetric ferroelectric tunnel junctions is studied using a multiscale thermodynamic
model. It is demonstrated in details that there exists a critical thickness at which
an unusual ferroelectric-\lq\lq polar non-ferroelectric\rq\rq phase transition occurs
in asymmetric ferroelectric tunnel junctions. In the \lq\lq polar non-ferroelectric\rq\rq
phase, there is only one non-switchable polarization which is caused by the competition
between the depolarizing field and the built-in field, and closure-like domains are
proposed to form to minimize the system energy. The transition temperature is found
to decrease monotonically as the ferroelectric barrier thickness is decreased and the
reduction becomes more significant for the thinner ferroelectric layers. As a matter of
fact, the built-in electric field does not only result in smearing of phase transition
but also forces the transition to take place at a reduced temperature. Such findings
may impose a fundamental limit on the work temperature and thus should be further taken
into account in the future ferroelectric tunnel junction-type or ferroelectric capacitor-type devices.
\end{abstract}

\maketitle

\section{INTRODUCTION}
Ferroelectric (FE) tunnel junctions (FTJs) that are composed of FE thin films of a few
unit cells sandwiched between two electrodes (in most cases the top and bottom electrodes
are different) have attracted much more attention during the last decade.~\cite{Tsymbal,KOHLSTEDT,Zhuravlev}
It is generally believed that the interplay between ferroelectricity and quantum-mechanical
tunneling plays a key role in determining tunnel electroresistance (TER) or tunneling
current and TER effect usually takes place upon polarization reversal. Due to the strong
coupling of FE polarization and the applied field, the electric-field control of
TER or tunneling current,~\cite{Tsymbal,KOHLSTEDT,Zhuravlev,Garcia0,Garcia1,Garcia2,VELEV,Bilc,PANTEL,Gruverman,Chanthbouala}
spin polarization,~\cite{Zhuravlev1,Duan,Sahoo,Duan1,Niranjan,Burton,VELEV1,Zhuravlev2,Garcia3,Hambe,Valencia,Meyerheim,Bocher,Lu1,PANTEL1} and electrocaloric effect~\cite{Liu} can be achieved, which makes FEs promising candidates for nondestructive FE storage,~\cite{Tsymbal,KOHLSTEDT,Zhuravlev,Garcia0,Garcia1,Garcia2,VELEV,Bilc,PANTEL,Gruverman}
FE memristor,~\cite{Chanthbouala} spintronics (magnetization),~\cite{Zhuravlev1,Duan,Sahoo,Duan1,Niranjan,Burton,VELEV1,Zhuravlev2,Garcia3,Hambe,Valencia,Meyerheim,Bocher,Lu1,PANTEL1}
or electrocaloric~\cite{Liu} devices. Meanwhile, another mechanically (including strain or
strain gradient) induced TER is found recently, which also shows their potential applications
in mechanical sensors, transducers and low-energy archive data storage decices.~\cite{Luo}
Note that having different electrodes for the FTJs (some experiments use conductive
atomic force microscope tips instead of the top electrodes) is usually required for a
large effect at low bias voltage though the FTJs with same electrodes may also display
interesting performances.~\cite{KOHLSTEDT,VELEV,VELEV1,Bilc} Also note that all the
functionalities in these devices are strongly related to the thermodynamic stability
and switching ability of FTJs.~\cite{Tsymbal,KOHLSTEDT,Zhuravlev,Garcia0,Garcia1,Garcia2,VELEV,VELEV1,Bilc,PANTEL,Gruverman,Garcia3,Hambe,PANTEL1,Luo,Chanthbouala,Zhuravlev1,Duan,Sahoo,Duan1,Niranjan,Burton,VELEV1,Zhuravlev2,Garcia3,Hambe,Valencia,Meyerheim,Bocher,Lu1,PANTEL1,Liu}
Therefore, a fundamental understanding of ferroelectricity of FTJs, especially their size
effects, is crucial at the current stage of research.

Unfortunately, no consensus has been achieved on whether there exists a critical thickness
$h_{c}$ below which the ferroelectricity disappears in FTJs, especially for those with
different top/bottom electrodes. It is believed that an electrostatic depolarizing field
caused by dipoles at the FE-metal interfaces is responsible for the size effect.~\cite{Mehta,Junquera,Kim,Pertsev,Gerra0,Tagantsev}
However, recent theoretical studies suggest that the choice of electrode material may lead
to smearing of size effect or even vanishing of $h_{c}$.~\cite{Stengel,Zheng0,Zheng1,Cai0,Cai1}
For example, it was reported that choosing Pt as electrodes would induce a strong interfacial
enhancement of the ferroelectricity in Pt/BaTiO$_{3}$(BTO)/Pt FTJs, where $h_{c}$ is only
0.08 BTO unit cell.~\cite{Stengel} In addition, the results of a modified thermodynamic
model~\cite{Zheng0,Zheng1} and first-principles calculations~\cite{Cai0,Cai1} both indicate
that BTO barrier with dissimilar electrodes, i.e. Pt and SrRuO$_{3}$ (SRO) electrodes, might
be free of deleterious size effects. In contrast, it has been reported that asymmetric combination
of the electrodes (including the same electrodes with different terminations) will result in
the destabilization of one polarization state making the asymmetric FTJs non-FE.~\cite{Gerra0,UMENO}
And the up-to-date studies reported that the fixed interface dipoles near the FE/electrode
interface is considered the main reason for that detrimental effect.~\cite{LIU,LU} Considering
the importance of the physics in FTJs with dissimilar top and bottom electrodes, we are strongly
motivated to investigate the size effect in such asymmetric FTJs.

It was pointed out as early as 1963 that the contribution of different electronic and chemical
environments of the asymmetric electrode/FE interfaces would induce a large long-range electrostatic
built-in electric field $\vec{E}_{bi}$ in FE thin films.~\cite{Simmons} $\vec{E}_{bi}$ becomes
more significant in asymmetric FTJs and should be taken in to account.~\cite{Gerra0,Tagantsev}
In this study, we use a multiscale thermodynamic model~\cite{Liu,Gerra0,Tagantsev} to investigate the
effect of such built-in electric field on the phase transition of asymmetric FTJs by neglecting
the short-range interface dipoles. As a result, we discover an unusual FE-\lq\lq polar non-FE\rq\rq
phase transition in asymmetric FTJs. Then, we make detailed analysis of the contribution of the
built-in electric field to FE phase transition, i.e. $h_{c}$, what happens below $h_{c}$, transition
temperature $T_{c}$, and temperature dependence of dielectric response of the asymmetric FTJs.

\section{MULTISCALE THERMODYNAMIC MODEL FOR THE FTJS}
\begin{figure}[h]
\includegraphics[width=8.5cm]{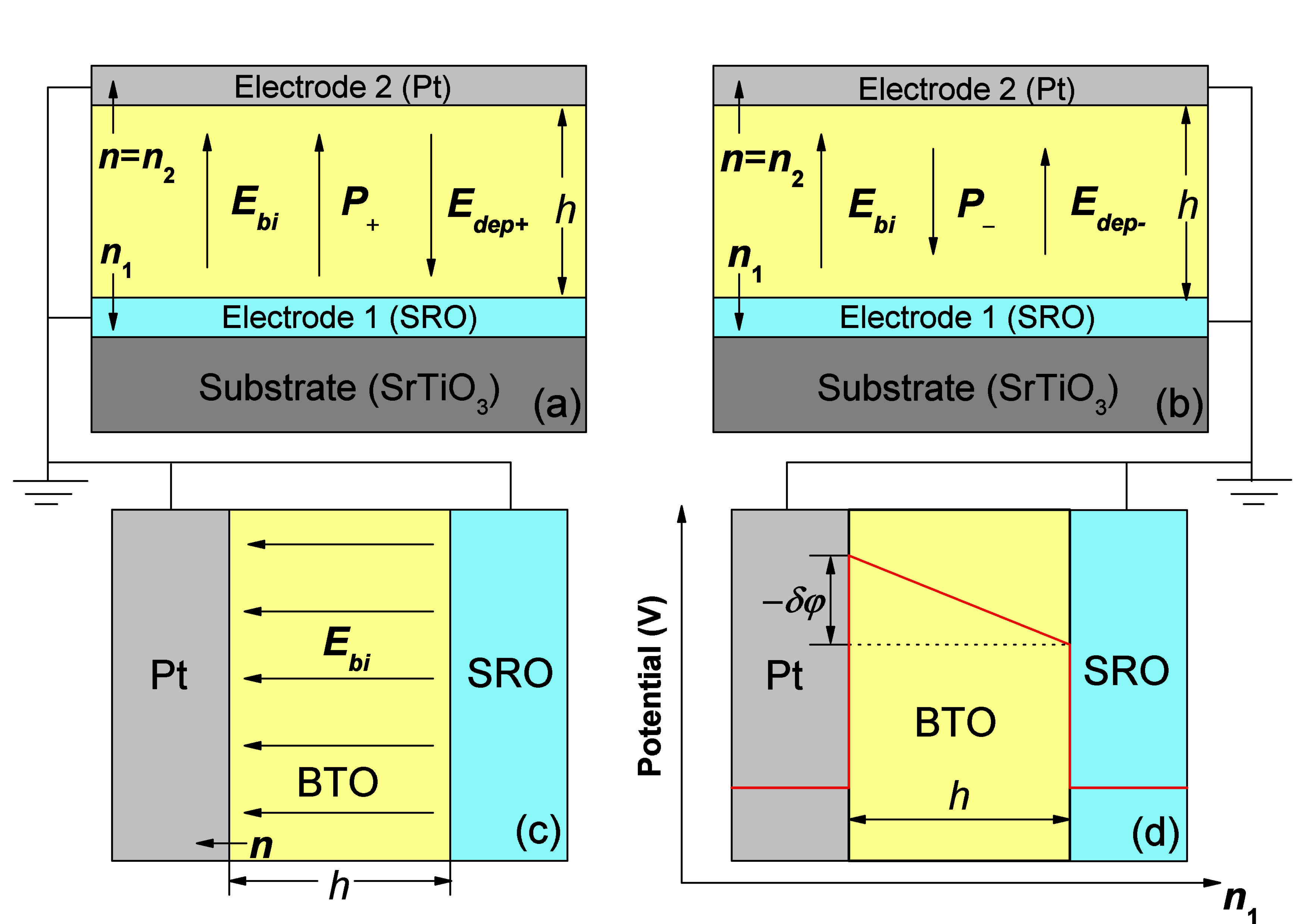}
  \caption{Schematic configurations of the system considered in the present
           calculations for the the asymmetric FTJs: $P_{+}$ state (a); $P_{-}$ state (b);
           $\vec{E}_{bi}$ (c) and the corresponding potential profile at zero polarization (red line) (d).}
  \label{Fig1}
\end{figure}

We concentrate on a short-circuited (001) single-domain FE plate of thickness $h$ sandwiched
between different electrodes. The FE films are fully strained and grown on thick (001) substrate
with the polar axis lying normal to the FE-electrode interfaces.~\cite{Pertsev,Gerra0,Tagantsev}
We denote the two interfaces as 1 and 2, with surface normals $\vec{n}_{1}$ and $\vec{n}_{2}=-\vec{n}_{1}$
pointing into the electrodes. The configurations are schematically shown in Fig.~\ref{Fig1}.
The exact value and direction of $\vec{E}_{bi}$ can be determined as~\cite{Gerra0,Tagantsev}
\begin{eqnarray}
\label{1}\vec{E}_{bi}=-\frac{\Delta\varphi_{2}-\Delta\varphi_{1}}{h}\vec{n}=-\frac{\delta\varphi}{h}\vec{n}\text{ and }\vec{n}=\vec{n}_{2}=-\vec{n}_{1},
\end{eqnarray}
where $\Delta\varphi_{i}$ which is the work function steps for FE-electrode $i$ interface
at zero polarization is simply defined as the potential difference between the FE and the
electrode $i$.~\cite{Gerra0,Tagantsev} With the help of first-principles calculations, one
could easily obtain $\Delta\varphi_{i}$ through the analysis of the electrostatic potential
of FTJs where FE films are in the paraelectric (PE) state.~\cite{Gerra0,Tagantsev}

Then, the free energy per unit surface of the FE layer is presented as~\cite{Gerra0,Tagantsev}
\begin{eqnarray}
\label{2}F&=&h\Phi+\Phi_{S}=(\frac{1}{2}\alpha_{1}^{*}P^{2}+\frac{1}{4}\alpha_{11}^{*}P^{4}+\frac{1}{6}\alpha_{111}P^{6}\nonumber\\
&+&\frac{1}{8}\alpha_{1111}P^{8}+\frac{u_{m}^{2}}{S_{11}+S_{12}}-\frac{1}{2}\vec{E}_{dep}\cdot\vec{P}-\vec{E}_{bi}\cdot\vec{P}\nonumber\\
&-&\vec{E}\cdot\vec{P})h+(\zeta_{1}-\zeta_{2})\vec{n}\cdot\vec{P}+\frac{1}{2}(\eta_{1}+\eta_{2})P^{2},
\end{eqnarray}
where $\alpha_{i}^{*}$ are Landau coefficients.~\cite{Liu} $u_{m}$ is the epitaxial strain
and $S_{mn}$ are the elastic compliances coefficients. $\zeta_{i}$ and $\eta_{i}$ are the
first order and second order coefficients of the surface energy $\Phi_{S}$ expansion for
the two FE-electrode interfaces.~\cite{Gerra0,Tagantsev} $\vec{E}$ is the applied electric
field along the polar axis. $\vec{E}_{dep}$ is the depolarizing field which can be determined
from the short-circuit condition such that:~\cite{Gerra0,Tagantsev}
\begin{eqnarray}
\label{3}\vec{E}_{dep}=-\frac{\lambda_{1}+\lambda_{2}}{h\varepsilon_{0}+(\lambda_{1}+\lambda_{2})\varepsilon_{b}}\vec{P},
\end{eqnarray}
where $\varepsilon_{0}$ is the permittivity of vacuum space, and $\varepsilon_{b}$ indicates
the background (i.e. without contribution of the spontaneous polarization) dielectric constant.
$\lambda_{i}$ are the effective screening lengths of the two interfaces and are dependent on
the polarization direction if the electronic and chemical environments of FE/electrode interfaces
are different.~\cite{Gerra0,Tagantsev} For the two opposite polarization orientations, the
direction dependence of $\lambda_{i}$ will induce the asymmetry in potential energy and hence
will produce the TER effect, besides the depolarizing field effect due to the polarization
difference between two opposite orientations.~\cite{Tsymbal,KOHLSTEDT,Zhuravlev} However, we
ignore such an effect due to the lack of information about the direction dependence of
$\lambda_{i}$ and we mainly focus on the role of the built-in field in this study. Note that
$\delta\varphi$ and $\delta\zeta=(\zeta_{2}-\zeta_{1})$ are thickness and polarization
independent and $\vec{E}_{bi}$ is indeed a long-range internal-bias field which has the effect
of poling the FE film.~\cite{Gerra0,Tagantsev,Simmons} In asymmetric FTJs, such asymmetry parameters
$\delta\varphi$ and $\delta\zeta$ can introduce a potential energy profile difference and
therefore induce the TER effect.~\cite{Tsymbal,KOHLSTEDT,Zhuravlev}

The equilibrium polarization can be derived from the condition of thermodynamic equilibrium:
\begin{eqnarray}
\label{4}\frac{\partial F}{\partial P}=0.
\end{eqnarray}

The dielectric constant $\varepsilon$ under an applied field $E$ whose direction is along the
polar axis can be determined as:~\cite{Zheng1}
\begin{eqnarray}
\label{5}\varepsilon=\frac{1}{h\varepsilon_{0}}(\frac{\partial^{2} F}{\partial^{2} P})^{-1}.
\end{eqnarray}

The multiscale thermodynamic model used in this study combines first-principles calculations and
phenomenological theory and its detailed description can be found elsewhere.~\cite{Gerra0,Tagantsev}
In the previous study, it is reported that $\vec{E}_{bi}$ could result in a smearing of the phase
transition and an internal-bias-induced piezoelectric response above $T_{c}$ in asymmetric FTJs.~\cite{Gerra0}
However, adding to the forgoing controversy on the size effects, further analysis of the effect
of built-in field on the FE transition in asymmetric FTJs is still absent. Inserting Eq. (\ref{1})
into Eq. (\ref{2}) results in a term that encompasses an odd power of the polarization:
$\vec{E}_{bi}\cdot\vec{P}$, which leads to asymmetric thermodynamic potentials. We shall show
that this term which behaves mathematically as identically as the phenomenological term suggested
by Bratkovsky and Levanyuk~\cite{Bratkovsky} will result in an unusual FE-\lq\lq polar non-FE\rq\rq
phase transition in asymmetric FTJs.

\section{RESULTS AND DISCUSSION}
\subsection{Size effects}
For a quantitative analysis, we consider a fully strained BTO film sandwiched between
Pt (electrode 2) and SRO (electrode 1) epitaxially grown on (001) SrTiO$_{3}$ substrates.
We neglect the energy difference of the asymmetric surfaces, i.e. by setting $\zeta_{1}=\zeta_{2}$,
$\eta_{1}=\eta_{2}$, to insure that the effect of $\vec{E}_{bi}$ is clearly observable
from the calculations since it is reported that surface effects are generally much smaller
than that of $\vec{E}_{bi}$.~\cite{Zheng0,Zheng1} All the parameters we used are listed
in Ref. 65. We first examine the effect of $\vec{E}_{bi}$ on the ferroelectricity of
asymmetric FTJs. Previous studies indicate that the direction of $\vec{E}_{bi}$ in asymmetric
Pt/BTO/SRO FTJs points to Pt electrode with higher work function.~\cite{Zheng0,Zheng1,Cai0}
All recent results show indeed that a strong preference for one polarization state namely $P_{+}$
while $P_{-}$ disappears at \lq\lq $h_{c}$\rq\rq.~\cite{Zheng0,Zheng1,Cai0,Cai1,UMENO,LIU,LU}
According to the definition of ferroelectricity, the spontaneous polarization of the FE
materials is switchable under an ac electric field.~\cite{Scott} However, knowing that
the spontaneous polarization of FE materials is switchable under an ac electric field,~\cite{Scott}
recent reports~\cite{Zheng0,Zheng1,Cai0,Cai1} are rather confusing and remain incomplete
on this point. Indeed, in addition to the forementioned divergence in the size effects,
two different transition temperatures at which the two polarization states reach zero
are obtained (see Ref. 37), which may be confusing since there should be only one finite
phase transition temperature for disappearance of ferroelectricity. In order to avoid such
confusions, we used the classical definition of ferroelectricicity~\cite{Scott} in the
following parts.

\begin{figure}[h]
\includegraphics[width=8.5cm]{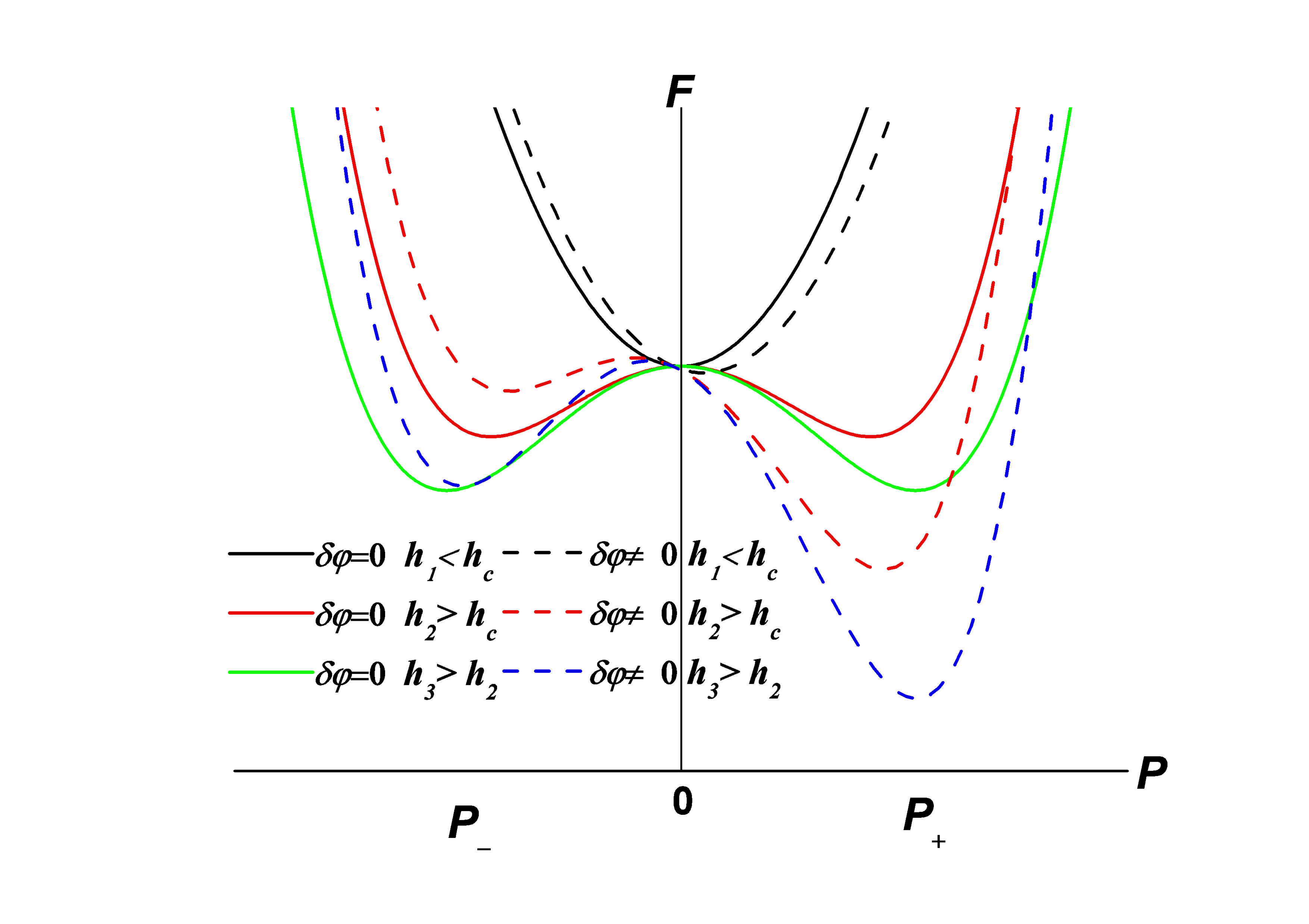}
  \caption{Schematic representation of the variation in total free energy with
           respect to polarization with different BTO barrier thicknesses in
           asymmetric Pt/BTO/SRO tunnel junctions with/without consideration
           of the built-in field in zero applied field.}
  \label{Fig2}
\end{figure}

We make further analysis of the physical formulation of $h_{c}$ in asymmetric FTJs. Note
that Eq. (\ref{4}) is a nonlinear equation and yields \lq\lq at most\rq\rq three solutions
$P$, two of them corresponding to minima and the other one to a saddle point
(unstable state). Whether the solution is a minimum, a maximum or a saddle point can be
revealed through inspecting the eigenvalues of the Hessian matrix of the total free energy
$F$. Because the asymmetric FTJ is internally biased, i.e., the energy degeneracy between
positive $P_{+}$ and negative $P_{-}$ is lifted, one of the minima corresponds to the
equilibrium state (the global minimum) of the system (the direction of which is along
$\vec{E}_{bi}$) and the other minima corresponds to a metastable state (a local minimum)
of the system. It means that the presence of two different electrodes in asymmetric FTJs
results in a preferred polarization orientation of the FE plate. Having found all $P$
solutions as a function of $h$, one can clearly see that metastable state and unstable
state solutions become closer to each other and coincide at finite $h$, henceforth the
number of solutions $P$ drops from three to one. According to the bistable property of FE
materials, this finite $h$ is just $h_{c}$.~\cite{Scott} As long as there are three $P$
solutions: two of these three solutions correspond to stable/metastable polarizations so that
two orientations of polarization are possible in the BTO layer and thus it is FE. Switching
the asymmetric FTJ into its unfavoured high energy polarization may be difficult. If there
is the only $P$ solution corresponding to the unstable state, although it attains a finite
value, it is not FE anymore and may be called \lq\lq PE\rq\rq. Indeed it would be more appropriate
to consider it as \lq\lq polar non-FE\rq\rq since $P$ has a unique finite value.~\cite{Liu1,Liu2,Okatan}
FTJs with no built-in field $\delta\varphi=0$ will exhibit two energetically equivalent stable
polarization states ($P_{+}$ and $P_{-}$) along with an unstable polarization state at $P=0$
below $h_{c}$. All the forgoing discussions can be clearly and easily understood in the
schematic representation of $F-P$ curves with different BTO thicknesses as shown in Fig.~\ref{Fig2}
which is quite similar with the results of FE thin films with/without consideration of the
fixed interface dipoles near the asymmetric FE/electrode interface~\cite{UMENO,LIU,LU} or FE
superlattices with/without interfacial space charges.~\cite{Okatan}
Together with previous results,~\cite{LIU,LU} we conclude that no matter the $\vec{E}_{bi}$
is considered as a long-range field or a short-range surface one, it cannot induce the
vanishing of $h_{c}$ in asymmetric FTJs, which is in contrast with other works.~\cite{Zheng0,Zheng1,Cai0,Cai1}

Note that \lq\lq polar non-FE\rq\rq phase is actually a pyroelectric phase because there is a
non-switchable polarization in this phase. This kind of phase transition has once been reported
in FE thin films with asymmetric electrodes~\cite{Gerra0,UMENO,LIU,LU} or FE superlattices
with interfacial space charge.~\cite{Liu1,Liu2,Okatan} As we discussed in the formation of $h_{c}$,
\lq\lq polar non-FE\rq\rq phase indeed always corresponds to the unstable state (see Fig.~\ref{Fig2})
and this kind of non-switchable polarization may not be stable at all. However, breaking up
the system into 180$^{0}$ domain stripes is unambiguously ruled out due to the long-range pinned
field $\vec{E}_{bi}$. In plane vortex formation~\cite{NAUMOV1,NAUMOV2} is also inhibited because
the large compressive strain favors more 180$^{0}$ domain stripes.~\cite{NAUMOV2} The
ferromagneticlike closure domains are predicted to form in ultrathin FE films or FE capacitors
even below $h_{c}$~\cite{Kornev,PUENTE,Shimada} and are experimentally confirmed well above $h_{c}$
recently.~\cite{Nelson,Jia} However, typical FE closure domains~\cite{Kornev,PUENTE,Shimada,Nelson,Jia}
are also not expected in \lq\lq polar non-FE\rq\rq phase where 180$^{0}$ domains in the closure
domain structure should be suppressed. But local rotations of non-switchable polarization ($<$90$^{0}$)
are still likely to occur and result in a closure-like domain structure since the local change
of the direction of the non-switchable polarization especially near the FE/electrode interface
is helpful to minimize the system energy.~\cite{Kornev,PUENTE,Shimada} Although such closure-like
domains can be favored below $h_{c}$ ($<$3 nm at least), it is clear that the FE barrier as a
whole is not FE according to our forgoing analysis that shows the polarization is not switchable under
external electric fields. While a detailed analysis of the built-in field effect on domain
formation is beyond the scope of this study, we suggest that more rigorous simulations should
be made in the future. It can be seen that the asymmetric FTJs below $h_{c}$ cannot be used
for FE memory applications in which two thermodynamic stable polarization states are needed
to encode \lq\lq 0\rq\rq and \lq\lq 1\rq\rq in Boolean algebra.~\cite{Mehta,Junquera,Kim,Pertsev,Gerra0,Tagantsev,Scott}
However based on our calculation, one should expect a resistance change below $h_{c}$ between
the non-switchable polarization state and the other one being ferroelectrically dead. This
result agrees well with recent works on Pt/BTO/Pt FTJs that even below $h_{c}$, the resistance
of the FTJ would change by a factor of three due to the interface bonding and barrier decay
rate effects.~\cite{VELEV} We argue that the TER effect below $h_{c}$ suggested in our work
may be essentially attributed to the asymmetric modification of the potential barrier by the
nonzero barrier height ($-\delta\varphi$) (see Eqs. (\ref{1})-(\ref{4})) which even exists
at zero polarization as shown in Fig.~\ref{Fig1}(d). Further theoretical and experimental
efforts should be made to confirm these predictions.
\begin{figure}[h]
\includegraphics[width=8.5cm]{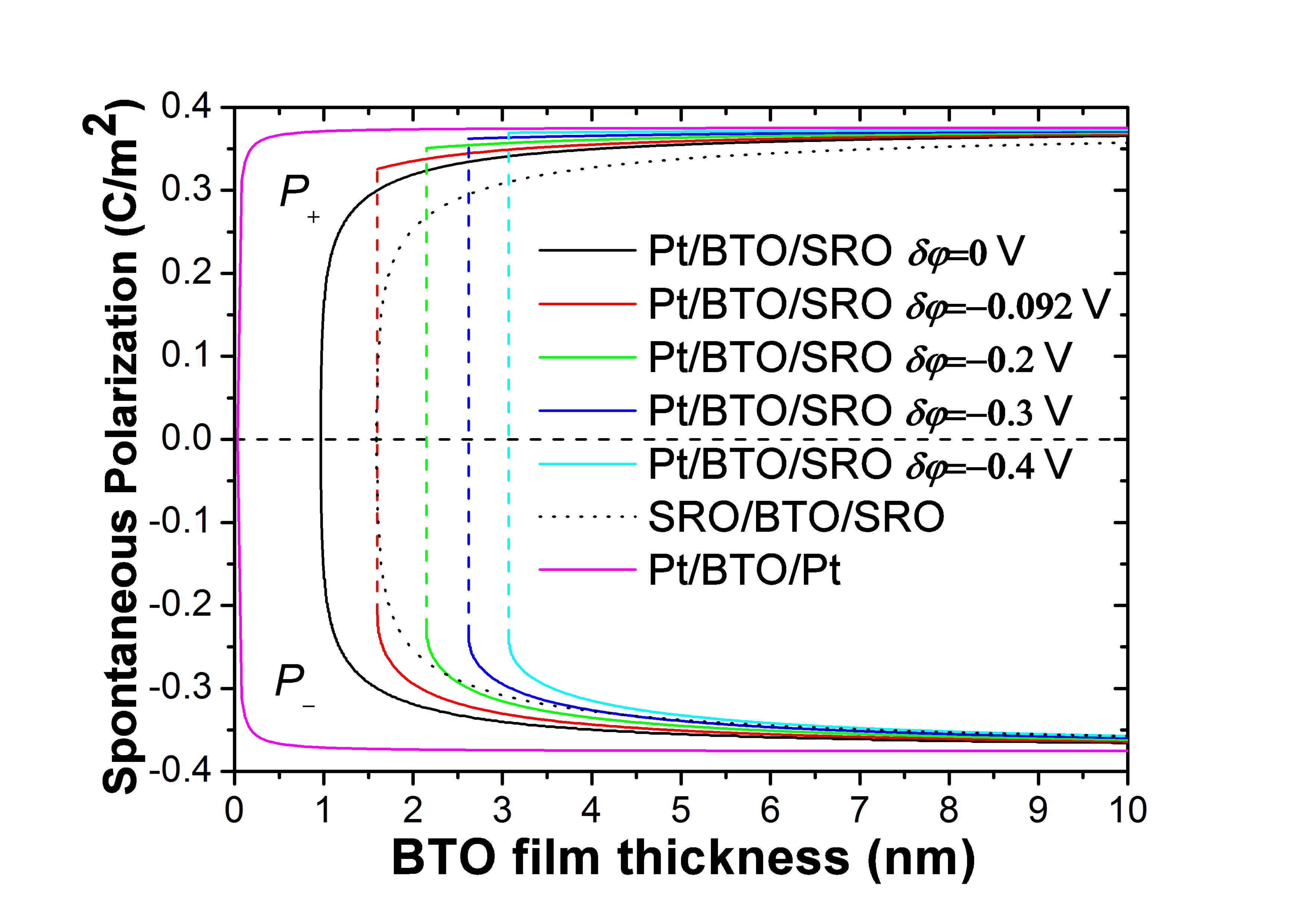}
  \caption{Spontaneous polarization of the asymmetric Pt/BTO/SRO tunnel junctions as
           a function of BTO layer thickness with $\delta\varphi$=0 V, -0.092 V, -0.2 V,
           -0.3 V and 0.4 V in zero applied field at 0 K, respectively. The results of
           symmetric SRO/BTO/SRO and Pt/BTO/Pt tunnel junctions at 0 K~\cite{Liu}
           are also added for comparison.}
  \label{Fig3}
\end{figure}

The quantitative results of the forgoing analysis are directly given in Fig.~\ref{Fig3}.
It can be seen that $h_{c}$ exists regardless of symmetric or asymmetric structures.
As expected, the curves of $P_{+}$ and $P_{-}$ are symmetric with respect to $P=0$ at
$\delta\varphi=0$ where $h_{c}$ is about 1 nm which is smaller than that of SRO/BTO/SRO,
i.e. 1.6 nm.~\cite{Liu,Stengel} When $\delta\varphi\neq0$, the supposed degeneracy
between $P_{+}$ and $P_{-}$ occurs, i.e. $P_{+}$ is enhanced while $P_{-}$ is reduced
so the coordinate of the center of the hysteresis loop along the polarization axis
[1/2($P_{+}$+ $P_{-}$)] is shifted along the direction of $P_{+}$. It is shown that
such a displacement of the hysteresis loop along the polarization axis becomes more
significant as the strength of $\vec{E}_{bi}$ increases. It may be attributed to the
imprint caused by $\vec{E}_{bi}$ such that the whole shape of the hysteresis loop will
shift along the direction of the field axis which is antiparallel to the direction of
$\vec{E}_{bi}$.~\cite{Scott} Besides, it is found that as $\delta\varphi$ increases,
$h_{c}$ increases, which indicates that $\vec{E}_{bi}$ can enhance the size of $h_{c}$.
Thus, whether $h_{c}$ of Pt/BTO/SRO junction is larger or smaller than that in the
SRO/BTO/SRO counterpart strongly depends on exact value of $\delta\varphi$ as shown
in Fig.~\ref{Fig3}.

For the symmetric structures (SRO/BTO/SRO and Pt/BTO/Pt FTJs), one can easily see in Fig.~\ref{Fig3}
that single domain in the FE layer destabilizes as the film thickness is decreased due
to the depolarizing field effect.~\cite{Liu,Mehta,Junquera,Kim,Pertsev,Gerra0,Tagantsev}
And it is shown in Fig.~\ref{Fig3} that Pt/BTO/Pt FTJ whose $h_{c}$ is merely 0.08 BTO unit
cell is nearly free of deleterious size effects,~\cite{Liu} which agrees well with the
result of first-principles calculations.~\cite{Stengel} $h_{c}$ of SRO/BTO/SRO FTJ is about
four BTO unit cells, which is consistent well with our previous results.~\cite{Liu} The
qualitative result that $h_{c}$ of Pt/BTO/Pt FTJ is smaller than that of SRO/BTO/SRO FTJ
in this work is consistent well with those of first-principles calculations~\cite{Stengel}
and lattice model.~\cite{Wang} However, our results are in contrast with previous works~\cite{Kim,Zheng0,Zheng1} predicting $h_{c}$ of SRO/BTO/SRO
FTJ to be smaller than that of Pt/BTO/Pt FTJ. In these previous works,~\cite{Zheng0,Zheng1}
Mehta \emph{et al}' electrostatic theory about the depolarizing field ($\vec{E}_{dep}=-\frac{\vec{P}}{\varepsilon_{b}}(1-\frac{h/\varepsilon_{b}}{l_{s1}/\varepsilon_{e1}+l_{s2}/\varepsilon_{e2}+h/\varepsilon_{b}})$ where $l_{s1}$ and $l_{s2}$ are Thomas-Fermi screening lengths and $\varepsilon_{e1}$ and $\varepsilon_{e2}$
are dielectric constants of electrode 1 and 2) is used~\cite{Mehta} while in our work we
used the \lq\lq effective screening length\rq\rq model to describe the depolarizing
field (see Eq. (\ref{3})). Note that we used the same parameters as Refs. 36 and 37 except
for the model of depolarizing field.~\cite{parameters} The distinct results are understandable
since it is generally accepted that imperfect screening should be characterized by effective
screening length (See Eq. (\ref{3})) rather than Thomas-Fermi one in Mehta \emph{et al}' model.~\cite{Junquera1}
In fact, the effective screening length at Pt/BTO interface is only 0.03 {\AA}~\cite{Stengel}
much smaller than that of Thomas-Fermi one $\sim$0.4 {\AA},~\cite{Kim} so a significantly
reduced depolarizing field is expected and it would result in nearly no $h_{c}$ in Pt/BTO/Pt
FTJs. Previous study attributes this freedom of size effects in the Pt/BTO/Pt structure to the
\lq\lq negative dead layer\rq\rq near the Pt/BTO interface,~\cite{Stengel} while we argue that
it may result directly from the fact that the effective screening length of Pt electrode is
extremely small since Bratkovsky and Levanyuk suggested the \lq\lq dead layer\rq\rq model is
totally equivalent as to consider an electrode with a finite screening length.~\cite{Bratkovsky1}
Here we ignore the effect of the extrinsic \lq\lq dead layer\rq\rq formed between metal
electrode (i.e. Au or Pt) and a perovskite FE (i.e. Pb(ZrTi)O$_{3}$ or BTO). Indeed, Lou and Wang found
that the \lq\lq dead layer\rq\rq between Pt and Pb(ZrTi)O$_{3}$ is extrinsic and could be removed
almost completely by doping 2\% Mn.~\cite{Lou} Experimentally, many researchers found that
SRO/BTO/SRO capacitors (as well as other perovskite FE structures with conductive oxide electrodes)
are free from passive layers.~\cite{Kim,Kim1,Jin} Recently, a very interesting experimental result
demonstrates that the RuO$_{2}$/BaO terminations at BTO/SRO interface, which is assumed
as many pinned interface dipoles and plays a detrimental role in stabilizing a switchable FE
polarization, can be overcome by depositing a very thin layer of SrTiO$_{3}$ between BTO layer
and SRO electrode.~\cite{LIU,LU} Nonetheless, it is still unclear whether such pinned interface dipoles
are intrinsic and can be found in other FE/electrode interfaces (i.e. SRO/PbTiO$_{3}$ and Pt/BTO).

\begin{figure}[h]
\includegraphics[width=8.5cm]{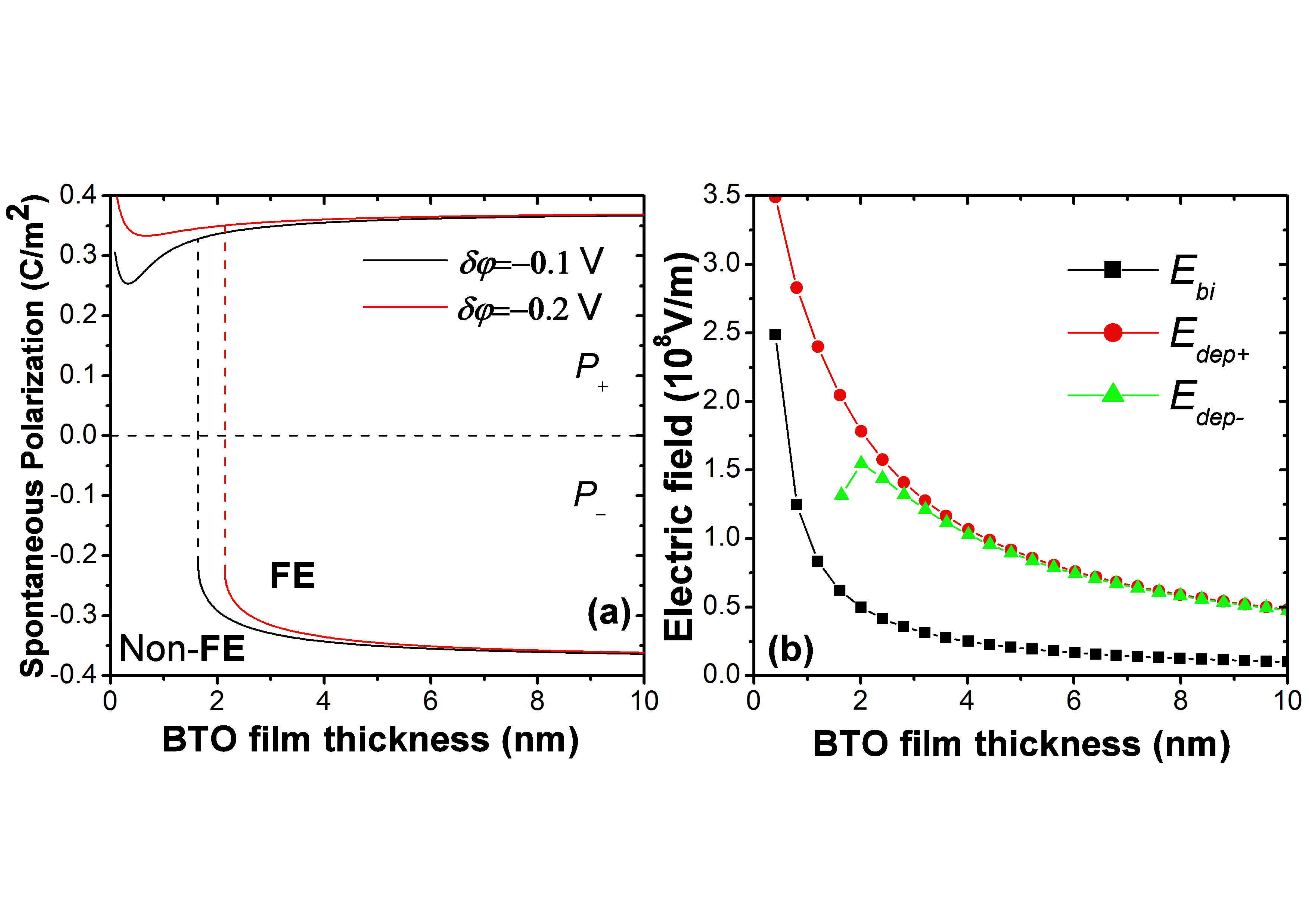}
  \caption{(a): Polarization state $P_{\pm}$ of the asymmetric Pt/BTO/SRO tunnel junctions
           as a function of BTO layer thickness with $\delta\varphi$=-0.1 V and -0.2 V
           in zero applied field at 0 K, respectively. The dash lines mark the boundary
           between polar non-FE and FE phases for different values of $\delta\varphi$;
           (b): Dependence of the strengths of the built-in field $E_{bi}$ and depolarizing
           field for different directions, $E_{dep+}$ and $E_{dep-}$ on the BTO layer
           thickness with $\delta\varphi$=-0.1 V at 0 K.}
  \label{Fig4}
\end{figure}

In the asymmetric structures in Fig.~\ref{Fig3}, it is shown that in comparison with
$\delta\varphi=0$ in asymmetric Pt/BTO/SRO FTJs, $h_{c}$ is significantly enhanced, as
$\delta\varphi$ increases, which is in good agreement with the recent results regarding
$\vec{E}_{bi}$ as short-range interface field,~\cite{LIU} and is similar with the previous
results.~\cite{Zheng1}  Note that $\delta\varphi$ is intrinsic and determined strictly by
the electronic and chemical environments of FE/electrode interfaces but not by any potential
drop through the FTJ which \lq\lq creates\rq\rq an applied field.~\cite{Gerra0,Tagantsev,Bratkovsky,Simmons}
Changing $\delta\varphi$ is simply due to the lack of its exact value and for the purpose
of studying the effect of $\vec{E}_{bi}$ in asymmetric FTJs, which is similar to the previous
method.~\cite{Zheng0,Zheng1} This method~\cite{Zheng0,Zheng1} indeed does not mean that any
asymmetric electrodes are considered here, since the electrode is replaced, the electrode/FE
interface parameters in Eqs. (\ref{1})-(\ref{3}) such as $\lambda_{i}$ and other interface
parameters will also change. The variation of $P_{+}$ in Pt/BTO/SRO FTJs as a function of
the whole BTO layer thickness with $\delta\varphi$=-0.1 V and -0.2 V at 0 K is shown in
Fig.~\ref{Fig4}(a). It is found that below the critical thickness $P_{+}$ state shows an
interesting recovery of a polar non-FE polarization, in contrast to $P_{-}$ state
(see Fig.~\ref{Fig4}(a)), becoming less significant when $\delta\varphi\sim<$-0.2 V.
Note that such recovery has been reported in FE superlattices with asymmetric electrodes
and demonstrated to be independent of the interfacial space charge.~\cite{Liu1} Although
such a recovery of polar non-FE polarization in BTO layer does not mean the recovery of
ferroelectricity as it is not switchable, it is necessary to realize its origin. We plot
the build-in field $E_{bi}$ and depolarizing field for different directions, $E_{dep+}$
and $E_{dep-}$, as a function of the BTO film thickness considering $\delta\varphi$=-0.1 V
as an example in Fig.~\ref{Fig4}(b). For the condition of $P_{-}$ state as schematically
illustrated in Fig.~\ref{Fig1}(b), $E_{dep-}$ shows the typical behavior as the FTJ with
the same electrodes,~\cite{Junquera,Kim,Pertsev,Wang} which means that $E_{dep-}$ plays a
key role forcing the single domain in the FE layer to destabilize as the film thickness is
decreased. $E_{bi}$ with the same direction of $E_{dep-}$ helps then to speed up such
destabilization, therefore enhancing the critical thickness. For the $P_{+}$ state,
$E_{dep+}$ and $E_{bi}$ are in the opposite directions, as depicted in Fig.~\ref{Fig1}(a),
and both the strengths of $E_{dep+}$ and $E_{bi}$ increase as the BTO layer thickness
is decreased (Fig.~\ref{Fig4}(b)), which means that $E_{dep+}$ is partially cancelled by
$E_{bi}$. The strength of this partial compensation becomes stronger with the film thickness
decreasing (see the slopes of $E_{bi}-h$ and $E_{dep+}-h$ curves)(Fig.~\ref{Fig4}(b)).
Therefore, $E_{bi}$ is fighting against $E_{dep+}$ allowing the polarization to recover
into a polar non-FE polarization. This recovery of polar non-FE polarization forces the
system to a higher energy state which strongly supports our forgoing predictions of
local rotations of non-switchable polarization ($<$90$^{0}$) and the formation of
closure-like domain structure to minimize the system energy.

\begin{figure}[h]
\includegraphics[width=8.5cm]{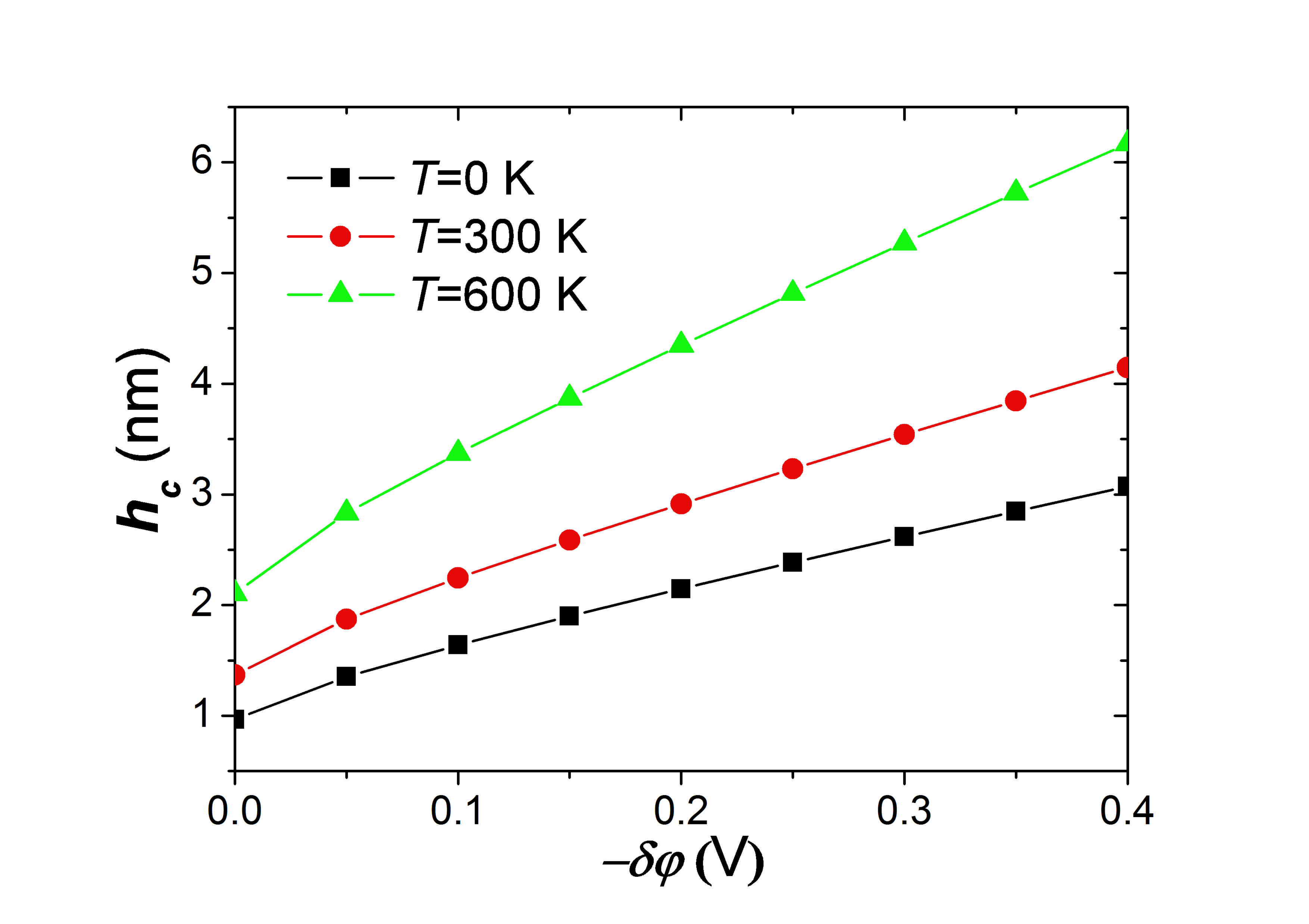}
  \caption{The variation in critical thickness $h_{c}$ in epitaxial asymmetric Pt/BTO/SRO
           tunnel junctions as a function of $-\delta\varphi$ at three different
           temperatures: 0 K, 300 K, and 600 K, respectively ($E$=0 kV/cm).}
  \label{Fig5}
\end{figure}

The critical thickness $h_{c}$ under different ambient temperatures $T$ as a function of
($-\delta\varphi$) in asymmetric Pt/BTO/SRO FTJs is shown in Fig.~\ref{Fig5}. It can be seen
that $h_{c}$ decreases with $T$ increasing. And it is found that for other $T$, the asymmetric
Pt/BTO/SRO FTJs show a similar behavior of enhancement of $h_{c}$ by increasing the strength
of $\vec{E}_{bi}$ as shown in Fig.~\ref{Fig3} at 0 K.

\subsection{Transition temperature and dielectric response}
\begin{figure}[h]
\includegraphics[width=8.5cm]{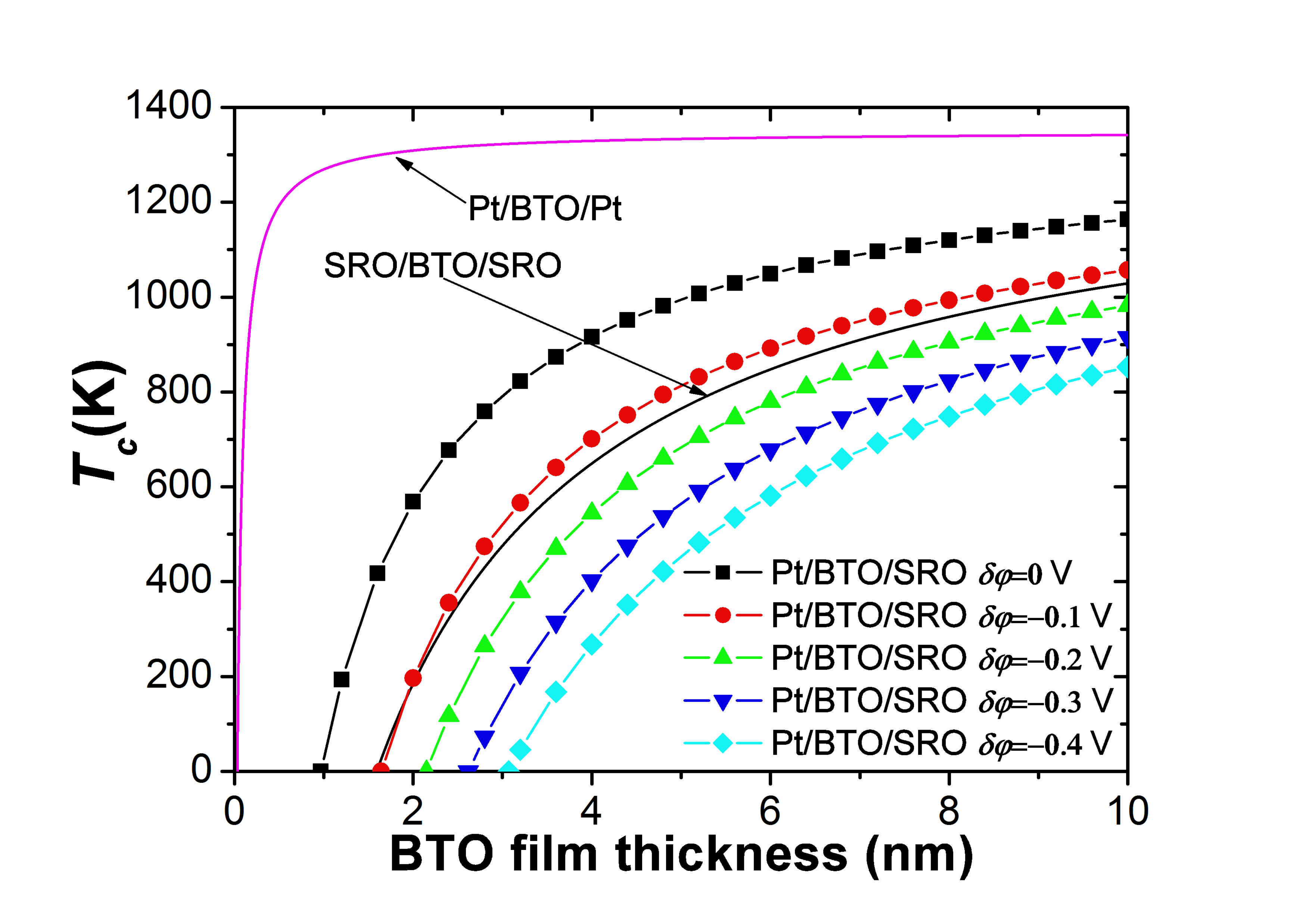}
  \caption{The transition temperature $T_{c}$ as a function of BTO layer thickness
           in epitaxial asymmetric Pt/BTO/SRO tunnel junctions at various values of
           $\delta\varphi$ with no applied field. The results of symmetric SRO/BTO/SRO
           and Pt/BTO/Pt tunnel junctions~\cite{Liu} are also provided for comparison.}
  \label{Fig6}
\end{figure}

\begin{figure}[h]
\includegraphics[width=8.5cm]{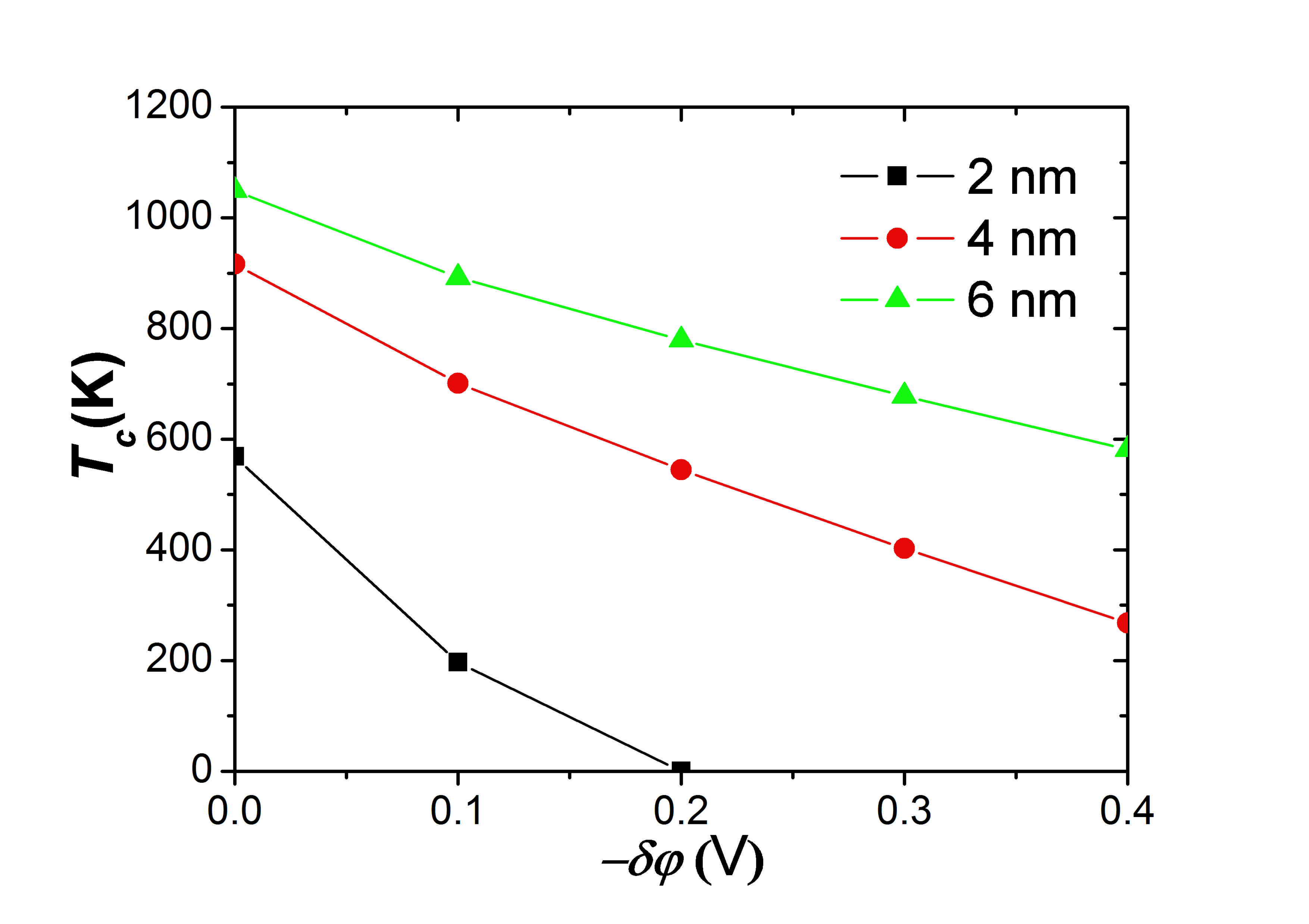}
  \caption{The transition temperature $T_{c}$ as a function of ($-\delta\varphi$)
           in epitaxial asymmetric Pt/BTO/SRO tunnel junctions with three different
           BTO layer thicknesses: 2 nm, 4 nm and 6 nm, respectively ($E$=0 kV/cm).}
  \label{Fig7}
\end{figure}

The transition temperature $T_{c}$ of the asymmetric FTJs is extremely important, especially
for the device applications. Fig.~\ref{Fig6} summarizes $T_{c}$ as a function of BTO thickness
in epitaxial asymmetric Pt/BTO/SRO FTJs at various values of $\delta\varphi$. It is shown that
$T_{c}$ in asymmetric Pt/BTO/SRO FTJs monotonically decreases with the BTO layer thickness
decreasing, which is similar to the behavior of symmetric SRO/BTO/SRO or Pt/BTO/Pt FTJs.~\cite{Liu}
Moreover, $T_{c}$ decreases more significantly for thinner BTO barrier layer thickness (see
the slope of $T_{c}-h$ curves in Fig.~\ref{Fig6}). At a given BTO layer thickness, it is found
in Fig.~\ref{Fig7} that $T_{c}$ decreases as $\delta\varphi$ becomes more negative, which
means a larger built-in field can force the phase transition to occur at lower temperatures.
The transition temperature $T_{c}$ is strongly sensitive to the $\delta\varphi$ change especially
for the thinner BTO barrier (see the slope of $T_{c}-(-\delta\varphi)$ curves in Fig.~\ref{Fig7}).
It can be clearly seen that the FE transition temperature is suppressed as the built-in field is
increased for different BTO thicknesses. Usually, the TER effect is always significantly larger
for thicker barrier with larger polarization.~\cite{Garcia0,Zhuravlev} Here we find that a
fundamental limit (which is more drastic for thinner FE barrier thickness) on the work temperature
of FTJ-type or capacitor-type devices should also be simultaneously taken into account together
with the FE barrier thickness or polarization value. In addition and interestingly, since the
electrocaloric effect is always the strongest close to the FE-PE transition,~\cite{Scott2} such
tuning of $T_{c}$ by $\vec{E}_{bi}$ should be also considered in potential asymmetric FTJs for
the room temperature solid-state refrigeration.~\cite{Liu} Moreover, the fact that large tunneling
current in asymmetric FTJs~\cite{Gruverman} results in significant Joule heating should also be
included in the design of future devices.

\begin{figure}[h]
\includegraphics[width=8.5cm]{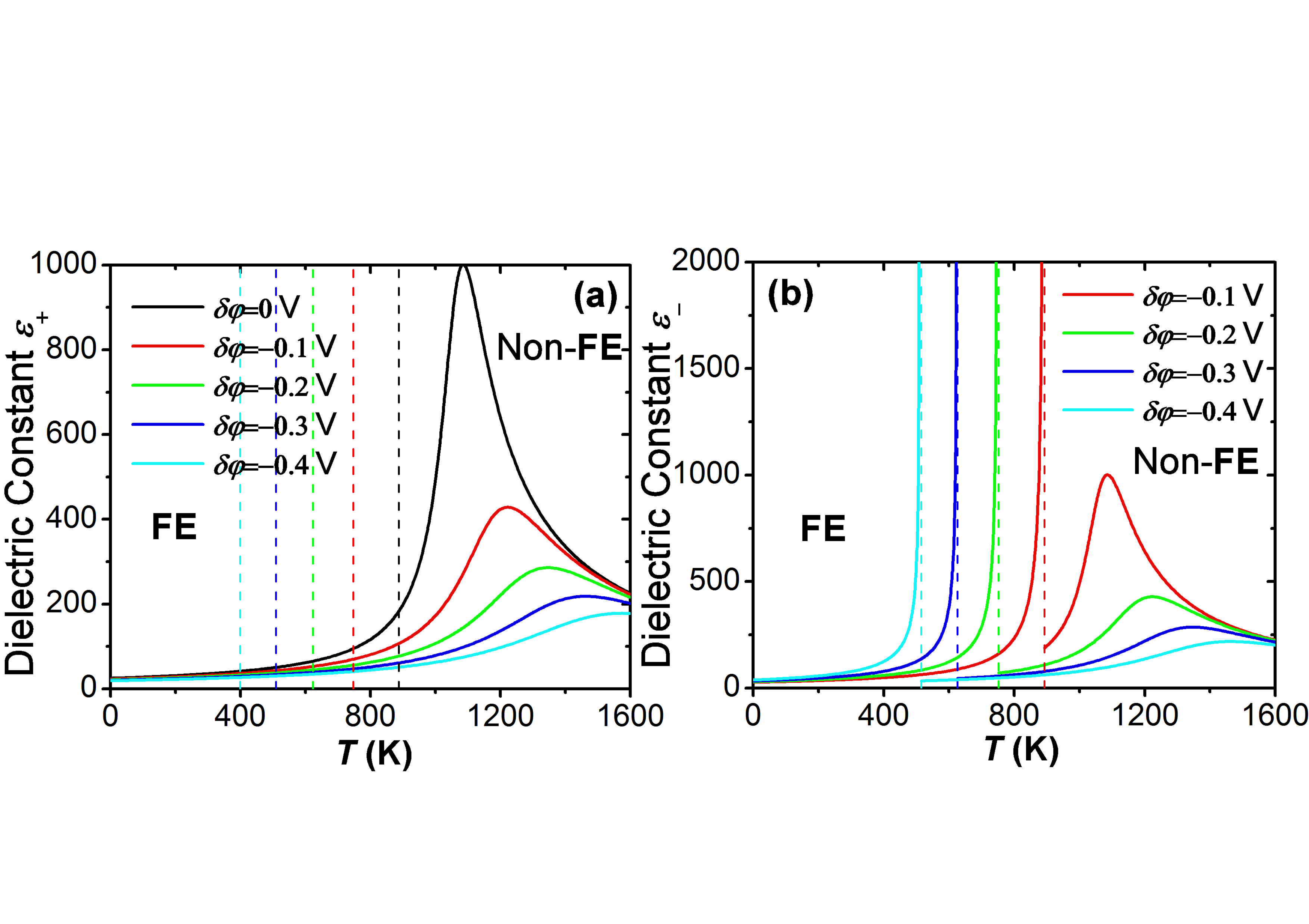}
  \caption{Dielectric constants $\varepsilon_{+}$ (a) and $\varepsilon_{-}$ (b)
           as a function of temperature $T$ at various values of $\delta\varphi$
           in asymmetric Pt/BTO/SRO tunnel junctions where BTO layer thickness
           $h$ is 5 nm ($E$=100 kV/cm). The dash lines mark the boundary between
           polar non-FE and FE phases for different values of $\delta\varphi$.}
  \label{Fig8}
\end{figure}

\begin{table}[h]
\caption{The different parameters extracted from Fig.~\ref{Fig8}(a).}
\begin{center}
\begin{tabular}{lllllllllll} 
\hline\hline $\delta\varphi$(V)& $T_{max}$(K)  & $\varepsilon_{max+}$ & $\varepsilon_{min+}$ & $\delta\varepsilon_{d}=(\varepsilon_{max+}-\varepsilon_{min+})/2$\\
\hline 0 \ \ & 1086\ \ & 1001.4\ \ &24.8\ \ &513.1\\
    -0.1 \ \ & 1223\ \ &428.5\ \ &23.0\ \ & 225.7\\
     -0.2 \ \ & 1345\ \ &285.8\ \ &21.4\ \ &153.6\\
     -0.3 \ \ & 1465\ \ &218.3\ \ &20.1\ \ &119.2\\
     -0.4 \ \ & 1574\ \ &178.1\ \ &18.9\ \ &98.5\\
\hline \hline
\end{tabular}
\end{center} \label{Fig8a parameters}
\end{table}

The dielectric response $\varepsilon_{+}$ ($\vec{E}$ is parallel to $\vec{E}_{bi}$) and
$\varepsilon_{-}$ ($\vec{E}$ is antiparallel to $\vec{E}_{bi}$) of Pt/BTO/SRO FTJs (consider
the 5-nm-thick BTO film as an example) as a function of $T$ at different $\delta\varphi$
is shown in Figs.~\ref{Fig8}(a) and (b). Several key parameters with different $\delta\varphi$
in Fig.~\ref{Fig8}(a) are extracted in Table.~\ref{Fig8a parameters}: $T_{max}$ corresponds
to the temperature where $\varepsilon_{+}$ reaches its maximum, $\varepsilon_{max+}$;
$\varepsilon_{min+}$ simply means the minimal value of $\varepsilon_{+}$; $\delta\varepsilon_{d}$
is in somehow the diffuseness of the transition. It can be seen that when $\delta\varphi=0$,
$\varepsilon_{+}$ shows a sharp peak near $T_{c}$. However, a gradual decrease in
$\varepsilon_{max}$ and $\delta\varepsilon_{d}$ is seen upon increasing $\varepsilon_{+}$
which is well consistent with the results of smearing of $T_{c}$ by increasing $\varepsilon_{+}$
in Fig.~\ref{Fig6} (see the slope of $T_{c}-h$ curves in Fig.~\ref{Fig6}). The diffusive
transition response in $\varepsilon_{+}$ clearly shows smearing of the phase transition as
a result of $\vec{E}_{bi}$, which verifies the predictions of Tagantsev \emph{et al}~\cite{Gerra0,Tagantsev}
and Bratkovsky \emph{et al}.~\cite{Bratkovsky} In addition, it is shown $T_{max}$ is shifted to
higher temperatures due to $\vec{E}_{bi}$. As the strength of $\vec{E}_{bi}$ increases, the
smearing of phase transition and the shift of $T_{max}$ becomes more significant. On the other
hand, the applied field cannot fully compensate the built-in field, resulting in a discountinuous
phase transition from FE phase to polar non-FE phase with temperature increasing as depicted in
dielectric response $\varepsilon_{-}$ in Fig.~\ref{Fig8}(b), which is distinct from the countinuous
counterpart of $\varepsilon_{+}$ as shown in Fig.~\ref{Fig8}(a). $P_{-}$ abruptly changes its
sign near the transition point resulting a dielectric peak and a similiar smearing of $\varepsilon_{-}$
by increasing the strength of $\vec{E}_{bi}$ is found. Furthermore, it is found that though the
transition temperatures for two directions are different, they both decrease as the built-in field
increases which is consistent with the results without any external field (See Fig.~\ref{Fig6}),
which indicates that the built-in field forces the transition to take place at a reduced
temperature.

\subsection{Comments on the built-in field effect}
We make further comments on the built-in field effect in asymmetric FTJs. The main assumption
in this study is that $\delta\varphi$ does not change during the polarization reversal.~\cite{Gerra0,Tagantsev}
The presence of $\delta\varphi$ which results in an asymmetric potential energy and barrier height
differences by switching the polarization will induce the TER effect.~\cite{Tsymbal,KOHLSTEDT,Zhuravlev}
Note that the switching of the polarization in the asymmetric FTJs may change the value of
$\delta\varphi$.~\cite{VELEV,VELEV1,UMENO,Stengel2,Chen} However, according to our analysis,
the variation in $\delta\varphi$ (even changing its sign occurs during the polarization reversal)
does not alter the main results of this study due to its induced broken spatial inversion symmetry
of FTJs. In addition to the built-in field, if the surface term $\delta\zeta=(\zeta_{2}-\zeta_{1})$
is nonzero, the main conclusions of this paper will not change as well.

\section{CONCLUSIONS}
In summary, on the basis of a multiscale thermodynamic model, a detailed analysis of the changes
brought by the built-in electric field in asymmetric FTJs is made. It is demonstrated that the
critical thickness does exist in asymmetric FTJs. Below the critical thickness, it is found
that there is a recovery of polar non-FE polarization due to strong cancelling of the depolarizing
field by the built-in field, and closure-like domains are proposed to form to minimize the system
energy. It is found that the built-in electric field could not only induce imprint and a behavior
of smearing of the FE phase transition but also forces the phase transition to take place at a
reduced temperature. A fundamental limit of transition temperature dependence of the barrier
layer thickness on the work temperature of FTJ-type or FE capacitor-type devices is proposed
and should be simultaneously taken into account in the further experiments. Hopefully, our
results will be helpful to the fundamental understandings of phase transitions in asymmetric FTJs.

\begin{acknowledgments}
This work is supported by the Ministry of Science and Technology of China through a 973-Project
under Grant No. 2012CB619401. The authors gratefully thank Dr. X. Y. Wang, Dr. M. B. Okatan,
and Prof. S. P. Alpay for their fruitful suggestions. Y. Liu is thankful to the Multidisciplinary
Materials Research Center (MMRC) at Xi'an Jiaotong University for hospitality during his visit.
Y. Liu and B. Dkhil wish to thank the China Scholarship Council (CSC) for funding YL's
stay in France. Y. Liu, M. Bibes and B. Dkhil also acknowledge the Agence Nationale
pour la Recherche for financial support through NOMILOPS (ANR-11-BS10-016-02) project.
X. J. Lou would like to thank the \lq\lq One Thousand Youth Talents\rq\rq program for support.
\end{acknowledgments}

\end{document}